\documentclass[twocolumn,aps,prb,superscriptaddress]{revtex4-2}
\usepackage{amsfonts}
\usepackage{amsmath}
\usepackage{amssymb}
\usepackage{graphicx}
\usepackage{dcolumn}
\usepackage{bm}
\usepackage{hyperref}
\usepackage{color}
\usepackage{siunitx}
\usepackage{mathrsfs}
\makeatletter
\hypersetup{hypertex=true, colorlinks=true, linkcolor=blue, anchorcolor=blue,citecolor=blue}

\begin{document}
\title{A generic model with unconventional Rashba bands and giant spin galvanic effect}
\author{Xinliang Huang}
\thanks{These authors make equal contributions.}
\affiliation{Anhui Province Key Laboratory of Low-Energy Quantum Materials and
		Devices, High Magnetic Field Laboratory, HFIPS, Chinese Academy
		of Sciences Hefei, Anhui, 230031, China}
\affiliation{Science Island Branch of Graduate School, University of Science and
		Technology of China, Hefei, Anhui 230026, China}
\author{Yuhang Xiao}
\thanks{These authors make equal contributions.}
\affiliation{Anhui Province Key Laboratory of Low-Energy Quantum Materials and
	Devices, High Magnetic Field Laboratory, HFIPS, Chinese Academy
	of Sciences Hefei, Anhui, 230031, China}
\affiliation{Science Island Branch of Graduate School, University of Science and
		Technology of China, Hefei, Anhui 230026, China}
\author{Rui Song}
\affiliation{Science and Technology on Surface Physics and Chemistry Laboratory,
		Mianyang, Sichuan 621908, China}
\affiliation{Anhui Province Key Laboratory of Low-Energy Quantum Materials and
	Devices, High Magnetic Field Laboratory, HFIPS, Chinese Academy
	of Sciences Hefei, Anhui, 230031, China}
\author{Ning Hao}
\email{haon@hmfl.ac.cn}
\affiliation{Anhui Province Key Laboratory of Low-Energy Quantum Materials and
	Devices, High Magnetic Field Laboratory, HFIPS, Chinese Academy
	of Sciences Hefei, Anhui, 230031, China}
	
\date{\today}
	
\begin{abstract}
  In two-dimensional system, Rashba spin-orbit coupling can lift spin degeneracy and gives the opposite spin chirality of two split Fermi circles from two Rashba bands. Here, we propose a generic model which can produce unconventional Rashba bands. In such a case, the two Fermi circles from two bands have the same spin chirality. When various interactions are taken into account, many unique physics can emerge in case of unconventional Rashba bands in comparison with in case of conventional Rashba bands. For instance, we study the spin galvanic effect by considering two cases with potential impurity scattering and magnetic impurity scattering, respectively. In both cases, we find the efficiency of spin galvanic effect is strongly enhanced in unconventional Rashba bands in comparison with conventional Rashba bands. More intriguingly, we find the effeiciency of conventional Rashba bands is insensitive to potential or magnetic impurity scattering. However, such  efficiency of uncoventional Rashba bands can be further enhanced by the magnetic impurity scattering in comparison with the potential impurity scattering. Thus, the unconventional Rashba bands can give giant spin galvanic effect. These results show that this model is useful to explore abnormal physics in the systems with unconventional Rashba bands.
	
\end{abstract}
\maketitle
	
\section{INTRODUCTION}
	The Rashba spin-orbit coupling (SOC) \cite{Rashba} is an interaction effect induced by relativistic SOC in systems with broken inversion symmetry \cite{bihlmayer2022rashba, manchon2015new, Bihlmayer_2015, koo2020rashba, WANG2024101145}. It elucidates the behavior of electrons at surfaces, interfaces, or heterostructures, leading to band splitting and spin-momentum locking, thereby forming chiral spin textures \cite{liu2019spintronic, PhysRevB.99.054407}. This effect has been extensively studied in various systems such as semiconductor heterostructures \cite{PhysRevLett.78.1335, chuang2015all, koo2009control, park2012observation, moharana2024chiralinduced, kang2024nanoscale}, metal surfaces or interfaces \cite{manchon2015new, PhysRevLett.98.186807, sanchez2013spin, hibino2021giant, PhysRevB.93.014420, PhysRevLett.117.116602, miron2011perpendicular, mihai2010current}, topological insulators \cite{PhysRevLett.116.096602, PhysRevB.97.174406, PhysRevB.94.184423, PhysRevB.96.235419, bahramy2012emergent, PhysRevLett.98.106803, RevModPhys.83.1057, RevModPhys.82.3045, PhysRevLett.120.037701, hoque2024room}, as well as artificial structures and low-dimensional materials \cite{mourik2012signatures, zhang2017ballistic, https://doi.org/10.1002/adma.202310768}. In the field of spintronics, it holds significant importance, providing a foundation for understanding and controlling spin-electronics and offering theoretical support for the design of novel spintronic devices and applications.
	
	Most systems with Rashba SOC exhibit band structures known as conventional Rashba bands, which feature spin textures of opposite chirality  for two Fermi circles at specified energy level. For instance, the spin texture observed on the Au(111) surface \cite{PhysRevLett.77.3419, PhysRevB.69.241401, PhysRevB.98.041404} described by the Bychkov-Rashba model \cite{Bychkov_1984} exemplifies this effect. Additionally, the Kane-Mele model \cite{PhysRevLett.95.146802,PhysRevLett.95.226801}, along with some extensions \cite{koo2020rashba, PhysRevB.90.165108, PhysRevB.82.161414, PhysRevB.94.085411, PhysRevLett.124.136403} incorporating Rashba SOC, also manifests the conventional Rashba effect. These models facilitate a deeper understanding of the conventional Rashba effect and provide guidance for related experiments.
	
	The unconventional Rashba bands are characterized by spin textures of the same chirality for two split Fermi circles with specified Fermi energy. Compared to the conventional Rashba bands, systems exhibiting unconventional Rashba bands are relatively rare. Few systems such as Bi/Cu(111), BiAg$_{2}$/Ag-Au(111) \cite{PhysRevB.79.245428, PhysRevB.94.041302, PhysRevB.100.115432, PhysRevLett.108.196801} and monolayer OsBi$_2$ \cite{PhysRevB.104.115433} are reported to show unconventional Rashba bands, and the mechanism is argued to originate from the between different bands and orbitals. However, the above systems are complicated due to the heterostructures and there still lacks a generic simple model to capture the key features of the unconventional Rashba bands. It sets block for us to explore the exotic physics in the systems with unconventional Rashba bands.
	
	In this work, we constructed a simplest four-band tight-binding (TB) model based on two-dimensional hexagonal and square lattices, with two orbitals degrees of freedom involved. By considering nearest-neighbor hopping, on-site SOC, and Rashba SOC, we demonstrate that the model adequately satisfies lattice symmetries and exhibits global unconventional Rashba bands, effectively describing systems with unconventional Rashba band structures. Remarkably, for unconventional Rashba bands near the $\Gamma$ point, one can yield a simple $\mathbf{k} \cdot \mathbf{p}$ model. This $\mathbf{k} \cdot \mathbf{p}$ model also features global unconventional Rashba bands, enabling simplification of the calculations.
	
	To explore the exotic physics implied in the unconventional Rashba band model, we investigate the inverse Edelstein effect (IEE) or spin galvanic effect \cite{bihlmayer2022rashba, RevModPhys.76.323, RevModPhys.87.1213, gomes2023surface, manchon2024spin} to demonstrate the advantages of this model in application on spintronics. This effect is about the conversion from spin current to charge current. The conversion efficiency is mainly governed by Rashba coefficients, the chirality of spin texture of involved Fermi circles and the momentum relaxation time \cite{mellnik2014spin, sanchez2013spin, ghiasi2019charge, cui2019field, he2021enhancement, PhysRevLett.123.207205, PhysRevApplied.12.034004, yu2021large}. For the Rashba coefficients, the tuning space is limited. For the chirality of spin texture, the advantage of the unconventional Rashba bands is apparent in comparison with the conventional Rashba bands. For the momentum relaxation time, which is usually determined by the impurity scattering in low temperature, we separately consider the impact from residual nonmagnetic and magnetic impurity scatting to it. For potential impurity scattering, we employed semi-classical Boltzmann transport theory to compute spin polarization, spin-to-charge conversion efficiency, and related quantities. We found that at low energy regimes, unconventional Rashba bands outperform conventional Rashba bands. As for magnetic impurity scattering, there are further complexities yet to be explored. To gain further insight into the scattering rate of electrons itinerating in the disordered circumstance containing random magnetic impurities, we go beyond the semi-classical treatment to calculate the second-order self-energy analytically. After carrying out the impurity-averaging procedure to recover translational symmetry, we obtain the Dyson equation of the impurity-averaged Green's function. Within the first-order Born approximation (FBA), we obtain the second-order Green's function explicitly and extract the scattering-rate information from it. All the transport quantities in the case of potential scattering are recalculated in this situation. We find that the conversion efficiency of uncoventional Rashba bands can be further enhanced by the magnetic impurity scattering in comparison with the potential impurity scattering. 
	
    This paper is organized as follows. First, we construct the unvonventional Rashba bands TB model for hexagonal, square lattices and further the $\mathbf{k} \cdot \mathbf{p}$ model in Sec. II. Next, we study the spin galvanic effect for both conventional and unconventional Rashba bands models through separately considering the nonmagnetic potential and magnetic impurity scatterings in Sec. III. At last, we give the conclusions in Sec. IV.   
    
\section{THE MODEL}
	
	In this section, we derive the TB model based on hexagonal and square lattices, as well as the corresponding $\mathbf{k} \cdot \mathbf{p}$ model. These models all have unconventional Rashba bands. Subsequently, for the $\mathbf{k} \cdot \mathbf{p}$ model, we will perform analytical calculations, and for the TB model, we will perform numerical calculations.
	
	\subsection{Hexagonal lattice}
	
\begin{figure*}[t]
		\includegraphics[width=0.9 \linewidth]{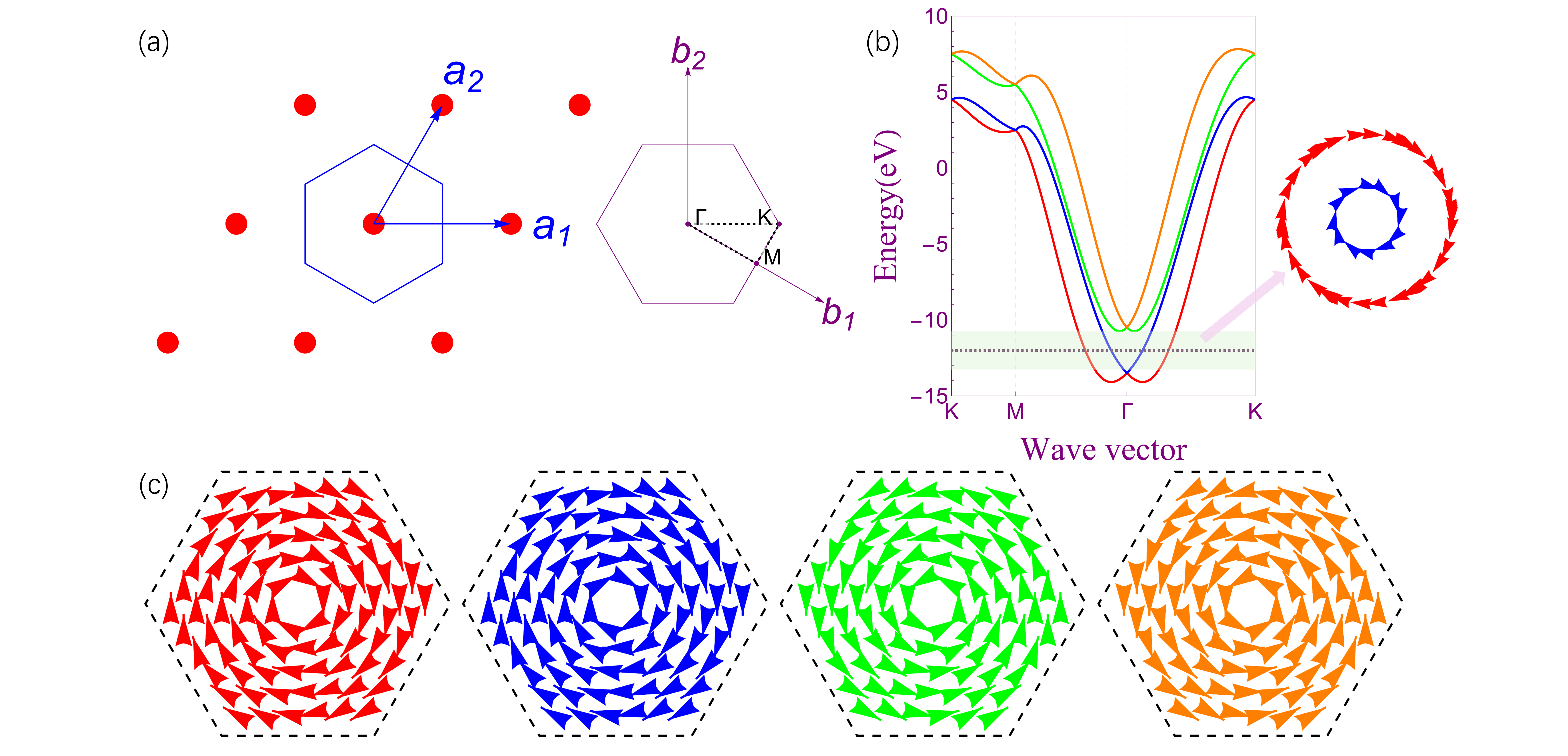}
		\caption{\label{fig:1}(a) A schematic diagram of the two-dimensional hexagonal lattice, with a direct lattice on the left, where $\mathbf{a}_1$ and $\mathbf{a}_2$ are direct lattice vectors, and a reciprocal lattice on the right, where $\mathbf{b}_1$ and $\mathbf{b}_2$ are reciprocal lattice vectors. (b) The band structure along a high symmetry path, where the plot on the right shows the spin texture at $-12\ \text{eV}$. (c) The global spin texture corresponding to each band corresponds one-to-one with the band color in (b). For (b) and (c), the parameters are set to $t=-2\ \text{eV},\lambda=3\ \text{eV},\lambda_R=0.7\ \text{eV} \text{\AA}$, respectively.}
	\end{figure*}

	A TB model is based on the two-dimensional hexagonal lattice shown in Fig.\ref{fig:1}(a), considering the hopping of the central atom to the surrounding six atoms. For each atom, we consider $d_{x^2-y^2}$ and $d_{xy}$ orbitals. Due to the breaking of inversion symmetry and the appearance of the Rashba effect, we specify that Fig.\ref{fig:1}(a) needs to satisfy the symmetry of the $C_{3v}$ group, which has six symmetric operations that can be labeled as $\{ E,2C_3,3\sigma_v \}$. The base of the model is $\Phi^\dag=(c_{d_{x^2-y^2},\uparrow}^\dag, c_{d_{x^2-y^2},\downarrow}^\dag, c_{d_{xy},\uparrow}^\dag, c_{d_{xy},\downarrow}^\dag)$, and the corresponding Hamiltonian is expressed as
	\begin{align}
		H=H_0 \otimes s_{0} + H_{\text{on-site SOC}} + H_{\text{Rashba SOC}} 
	\end{align}
	where $s_{0}$ is a $2 \times 2$ identity matrix, $H_0$ represents the Hamiltonian without considering SOC, which is the nearest neighbor hopping term, $ H_{\text{on-site SOC}}$ is the on-site SOC term, and $H_{\text{Rashba SOC}}$ is the nearest neighbor Rashba SOC term.
	
	For the Hamiltonian without considering SOC, we write it as
	\begin{align}
		H_0=\sum_i \sum_{\alpha} \varepsilon_\alpha c_{i,\alpha}^\dagger c_{i,\alpha} + \sum_{i,j} \sum_{\alpha,\beta} t_{\alpha \beta}^{ij} c_{i,\alpha}^\dagger c_{i,\alpha} 
	\end{align}
	where $i,j$ represents the primitive cell index, and $\alpha,\beta$ represents the orbital index. Since we are considering the case of two orbitals on an atom, in this case $i,j=1$ and $\alpha,\beta=d_{x^2-y^2}\ or \ d_{xy}$. Through Fig.\ref{fig:1}(a), we can see that the hopping from the central atom to the six surrounding atoms can be interconnected through $C_{3}$ and $\sigma_v$ operations, so we denote the hopping parameter as $t$. At the same time, in order to make the model as simple as possible, we set $\varepsilon_\alpha$ to 0 because it only changes the Fermi level (chemical potential), and ignore the coupling between $d_{x^2-y^2}$ and $d_{xy}$ orbitals. Thus, $H_0$ is represented as
	\begin{equation}{\tiny }   \label{H0}
		H_0=\begin{pmatrix}
			h_{0} & 0 \\
			0 & h_{0}
		\end{pmatrix}
	\end{equation}
	where
	\begin{gather}
		h_{0}=2 t \left(2 \cos \xi \cos \zeta + \cos 2\xi \right) \\
		\left( \xi,\zeta \right)=\left( \frac{k_x}{2},\frac{\sqrt{3} k_y}{2} \right)
	\end{gather}
	
	The on-site SOC term written as
	\begin{align}
		H_{\text{on-site SOC}} = \frac{\lambda}{2} \mathbf{S}\cdot \mathbf{L}
	\end{align}
	where $\mathbf{L}=(L_x,L_y,L_z)$ is the angular-momentum operator, and $\mathbf{S}=(s_x,s_y,s_z)$  is the spin operator, i.e.,  the Pauli matrix. The on-site SOC term is related to the selection of the base. According to our basis, where the orbitals are $d_{x^2-y^2}$ and $d_{xy}$, $H_{on-site\ SOC}$ is represented as
	\begin{equation}
		H_{\text{on-site SOC}}=\begin{pmatrix}
			0 & 0 & -i\frac{\lambda}{2} & 0\\
			0 & 0 & 0 & i\frac{\lambda}{2}\\
			i\frac{\lambda}{2} & 0 & 0 & 0\\
			0 & -i\frac{\lambda}{2} & 0 & 0
		\end{pmatrix}
	\end{equation}
	The on-site SOC mentioned here refers to a type of localized SOC, describing the interaction between the electron's spin motion and its orbital motion within a single atomic site, often resulting in the splitting of atomic energy levels. In our model, this lead to the splitting of the band structure at the $\Gamma$ point.
	
	The nearest neighbor Rashba SOC term breaks the inversion symmetry, resulting in spin textures in the system. The Rashba SOC formula is
	\begin{equation}
		h_{\text{Rashba SOC}}=i\lambda_R\sum_{i,j}c_i^\dag (\boldsymbol{s}\times \hat{\boldsymbol{d}}_{ij})_z c_j
	\end{equation}
	where $\boldsymbol{s}=(s_x, s_y, s_z)$ is the Pauli matrix, and $d_{ij}$ is the vector between lattice points, so there is
	\begin{align}
		\boldsymbol{s}\times \hat{\boldsymbol{d}}_{ij}=s_x(\hat{d}_{ij})_y-s_y(\hat{d}_{ij})_x
	\end{align}
	We consider the nearest neighbor hopping and obtain the Hamiltonian as
	\begin{align} \label{eq9}
			h_{\text{Rashba SOC}}=\begin{pmatrix}
				0 & \square \\
				\dag & 0
			\end{pmatrix}
	\end{align}
	where
	\begin{align}
		\square = -2 i \lambda _R\left(\sin \xi \cos \zeta - i \sqrt{3} \cos \xi \sin \zeta + \sin 2\xi \right)
	\end{align}
	The $h_{\text{Rashba SOC}}$ represents the Rashba SOC term between spin up and spin down. Our model comprises two orbitals, each considering spin up and spin down. That is to say, we consider Rashba SOC between different orbitals and within the same orbital. Based on the basis we use, the $H_{\text{Rashba SOC}}$ is represented as
	\begin{align}
		H_{\text{Rashba SOC}} = s_x \otimes h_{\text{Rashba SOC}} + s_0 \otimes h_{\text{Rashba SOC}}
	\end{align}
	where $s_{0}$ is a $2 \times 2$ identity matrix, $s_{x}$ is the $x$-component of the Pauli matrix. Note that, $s_x \otimes h_{\text{Rashba SOC}}$ represents the Rashba SOC term between different orbitals, while $s_0 \otimes h_{\text{Rashba SOC}}$ represents the Rashba SOC term within the same orbital.
	
	Compared to the Kane-Mele model \cite{PhysRevLett.95.146802,PhysRevLett.95.226801}, it can be observed that $H_{\text{Rashba SOC}}$ not only accounts for the coupling between spin up and spin down of different orbitals, but also between spin up and spin down of the same orbital. Additionally, after obtaining the individual Rashba SOC for spin up and spin down (denoted as $h_{\text{Rashba SOC}}$), we can further expand it to obtain the total Rashba SOC term (denoted as $H_{\text{Rashba SOC}}$). This expansion is arbitrary, and different materials may have different expansion methods. The basis we used is $\Phi^\dag=(c_{d_{x^2-y^2},\uparrow}^\dag, c_{d_{x^2-y^2},\downarrow}^\dag, c_{d_{xy},\uparrow}^\dag, c_{d_{xy},\downarrow}^\dag)$, which is equivalent to considering all Rashba SOC interactions between spin up and spin down after expansion.
	
	After determining each Hamiltonian matrix element, add the above terms together to obtain the Hamiltonian involving SOC as
	\begin{equation}{\tiny }  \label{h6}
		H=\begin{pmatrix}
			H_{11} & H_{12} & H_{13} & H_{14} \\
			& H_{22} & H_{23} & H_{24} \\
			& & H_{33} & H_{34} \\
			\multicolumn{2}{c}{\raisebox{1.3ex}[0pt]{\Huge \dag}}& & H_{44}
		\end{pmatrix}
	\end{equation}
	where
	\begin{gather}
			H_{11}=2 t \left(2 \cos \xi \cos \zeta + \cos 2\xi \right)\\
			H_{22}=H_{33}=H_{44}=H_{11}\\
			H_{12}=-2 i \lambda _R\left(\sin \xi \cos \zeta - i \sqrt{3} \cos \xi \sin \zeta + \sin 2\xi \right)\\
			H_{34}=H_{14}=H_{12}\\
			H_{23}=H_{12}^{*}\\
			H_{13}=-i\frac{\lambda}{2}\\
			H_{24}=H_{13}^{*}
	\end{gather}
	Here, we set $t=-2\ \text{eV},\lambda=3\ \text{eV},\lambda_R=0.7\ \text{eV} \text{\AA}$, and obtain the band structure along a high symmetry path, as well as the spin texture of each band, as shown in Fig.\ref{fig:1}(b) and Fig.\ref{fig:1}(c). Through the correspondence between Fig.\ref{fig:1}(b) and Fig.\ref{fig:1}(c), it can be observed that the coupled bands exhibit globally consistent spin textures of the same chirality. For the two coupled bands, such as the red and blue bands (or the green and orange bands), they possess the same chiral spin texture. At the specified energy level, the same chiral spin texture of the coupled band can always be obtained without worrying about chiral flipping. We hope to cut only a pair of spin textures with the same chirality in the first Brillouin zone, which is pure and robust. According to Fig.\ref{fig:1}(b), we should cut between the lowest point of the upper two coupled bands (i.e., the green and orange bands) and the intersection point of the lower two coupled bands (i.e., the red and blue bands). Here, we choose to cut at $E_F=-12 \ \text{eV}$, and we can see that the red and blue bands exhibit the same chiral spin textures, as shown in Fig.\ref{fig:1}(b). Subsequently, we will calculate the spin galvanic effect of the TB Model using the coupled bands (i.e., the red and blue bands).
	
	\subsection{Square lattice}
	
	\begin{figure*}[t]
		\includegraphics[width=0.9 \linewidth]{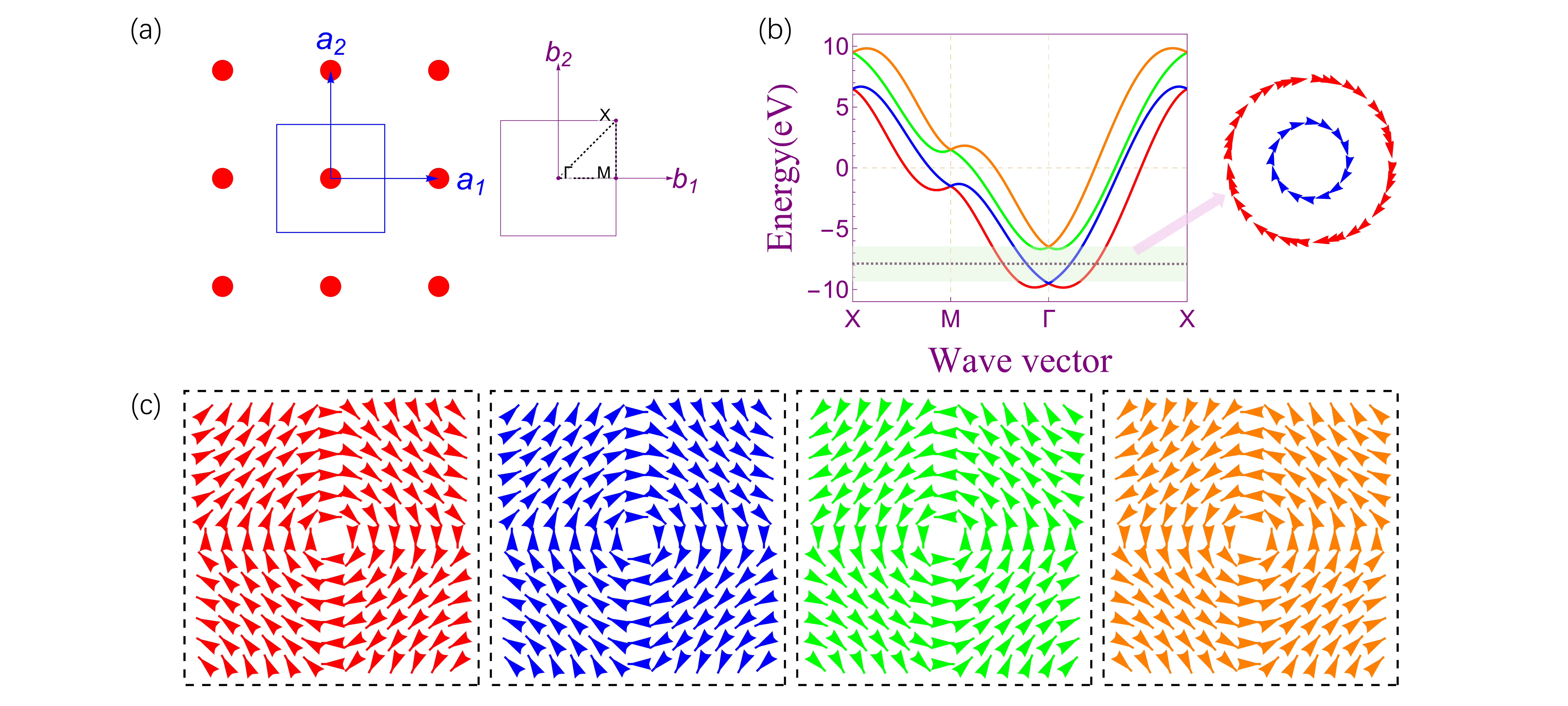}
		\caption{\label{fig:2}(a) A schematic diagram of the two-dimensional square lattice, with a direct lattice on the left, where $\mathbf{a}_1$ and $\mathbf{a}_2$ are direct lattice vectors, and a reciprocal lattice on the right, where $\mathbf{b}_1$ and $\mathbf{b}_2$ are reciprocal lattice vectors. (b) The band structure along a high symmetry path, where the plot on the right shows the spin texture at the $-8 \ \text{eV}$. (c) The global spin texture corresponding to each band corresponds one-to-one with the band color in (b). For (b) and (c), the parameters are set to $t=-2\ \text{eV},\lambda=3\ \text{eV},\lambda_R=0.7\ \text{eV} \text{\AA}$, respectively.}
	\end{figure*}
	
	Through the analysis above, similarly, we can also choose a square lattice to establish the TB model. The schematic diagram is shown in Fig.\ref{fig:2}(a). It can be observed that by adding the Rashba SOC term in the same manner, we can derive a TB model describing unconventional Rashba effects within the square lattice. The Hamiltonian here is very similar to that of the hexagonal lattice, with all other relations remaining unchanged; the only difference lies in the substitution as follows: 
	\begin{gather}
		H_{11} \rightarrow 2t(\cos \xi + \cos \zeta ) \\
		H_{12} \rightarrow -2\lambda_R \sin \zeta - 2i\lambda_R \sin \xi \\
		\left( \xi,\zeta \right) \rightarrow \left( {k_x},{k_y} \right)
	\end{gather}
	As before, we set $t=-2\ \text{eV},\lambda=3\ \text{eV},\lambda_R=0.7\ \text{eV} \text{\AA}$, and the relevant results are shown in Fig.\ref{fig:2}(b) and (c). The coupled bands feature globally consistent spin textures of the same chirality, i.e., unconventional Rashba bands.
	
	As a comparison, we can see that although the TB model of a square lattice is simpler than that of a hexagonal lattice, the physics given is the same. Of course, these two types of lattices are also the most common in two-dimensional materials. For a specific two-dimensional material, it is feasible to consider other interactions, such as the Zeeman field in magnetic systems, based on the existing TB model.
	
	\subsection{$\mathbf{k} \cdot \mathbf{p}$ model}
	
	We focus on the unconventional Rashba bands near the $\Gamma$ point in Fig.\ref{fig:1}(b) and Fig.\ref{fig:2}(b). To this end, we can perform a Taylor expansion of the TB model around the $\Gamma$ point to obtain a $\mathbf{k}\cdot\mathbf{p}$ model. This enables analytical solutions, facilitating the analysis of unconventional Rashba bands and the spin galvanic effect. For Eq.(\ref{h6}), we perform the Taylor expansion at the $\Gamma$ point to obtain
	\begin{equation}  \label{eq13}
		H_{\mathbf{k} \cdot \mathbf{p}}=\begin{pmatrix}
			t k^2 & -i\lambda_R k_{-} & -i\lambda & -i\lambda_R k_{-} \\
			& t k^2 & i\lambda_R k_{+} & i\lambda \\
			& & t k^2 & -i\lambda_R k_{-} \\
			\multicolumn{2}{c}{\raisebox{1.3ex}[0pt]{\Huge \dag}}& & t k^2
		\end{pmatrix}
	\end{equation}
	where $k^2=k_x^2+k_y^2$ and $k_{\pm}=k_x \pm ik_y$. Since the $\mathbf{k} \cdot \mathbf{p}$ model can be solved analytically, we apply the following formula to calculate the spin texture
	\begin{equation}  \label{eq14}
		\begin{split}
			S_k=\langle\Psi(k)|\Omega|\Psi(k)\rangle
		\end{split}
	\end{equation}
	where $\Omega=\tau_0 \otimes \mathbf{s}$, $\tau_0$ and $\mathbf{s}$ spans orbital space and spin space, respectively. The $\mathbf{k} \cdot \mathbf{p}$ model effectively captures the band information near the $\Gamma$ point in Fig.\ref{fig:1}(b) and Fig.\ref{fig:2}(b). As shown earlier, we analyze and solve the $\mathbf{k} \cdot \mathbf{p}$ model, and obtain the spin textures of the first and second bands (i.e., the red and blue bands) as follows:
	\begin{gather}
		(S_x^{1,2}, S_y^{1,2})=\frac{k\lambda_R}{\sqrt{k^2\lambda_R^2+\lambda^2}}(\sin\theta \boldsymbol{\hat{x}} - \cos\theta \boldsymbol{\hat{y}})
	\end{gather}
	The spin texture of the third and fourth bands (i.e., the green and orange bands) is
	\begin{gather}
		(S_x^{3,4}, S_y^{3,4})=\frac{k\lambda_R}{\sqrt{k^2\lambda_R^2+\lambda^2}}(-\sin\theta \boldsymbol{\hat{x}} + \cos\theta \boldsymbol{\hat{y}}) 
	\end{gather}
	where $k^2=k_x^2+k_y^2$ and $\theta=\arctan (k_y / k_x)$. It can be observed that the spin textures of the two coupled bands exhibit the same chirality. These results align well with the spin texture near the center (i.e., $\Gamma$ point) of Fig.\ref{fig:1}(c) and Fig.\ref{fig:2}(c), suggesting that the $\mathbf{k} \cdot \mathbf{p}$ model is suitable for subsequent calculations of unconventional Rashba bands.
	
\section{THE GIANT SPIN GALVANIC EFFECT}
	
\begin{figure}[t]
		\includegraphics[width=0.9 \linewidth]{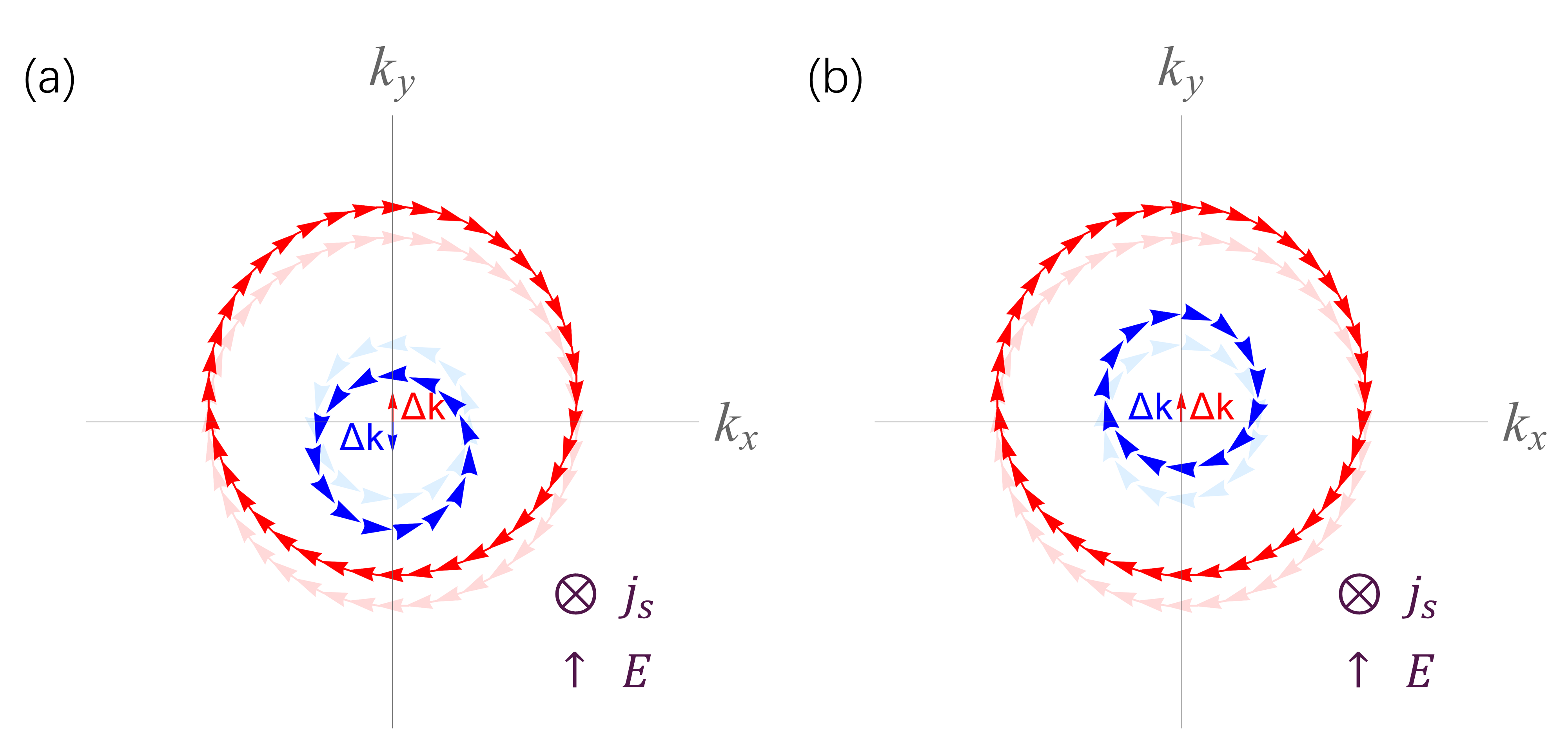}
		\caption{\label{fig:3}After injecting spin along the $\mathbf{z}$-direction or applying an electric field in the $\mathbf{y}$-direction, (a) and (b) correspond to the movement of the inner and outer circles of conventional and unconventional Rashba bands, respectively.}
	\end{figure}

	The spin galvanic effect, also known as the inverse Edelstein effect (IEE), discusses the conversion from spin current to charge current, with the corresponding conversion efficiency defined as $\lambda_{IEE}=j_c/j_s$, where $j_s$ is the pure spin current, and $j_c$ is the transverse charge current. This effect requires the injection of spin current or, equivalently, the application of an in-plane electric field. Fig.\ref{fig:3}(a) and Fig.\ref{fig:3}(b) illustrate schematics of the spin galvanic effect in conventional Rashba bands and unconventional Rashba bands, respectively.
	
	The model we propose features the unconventional Rashba bands, where the system is in equilibrium when spin is not injected or no electric field is applied, corresponding to the lightred and lightblue spin textures in Fig.\ref{fig:3}(b). After injecting spin or applying an electric field, the system exhibits the IEE, which is a non-equilibrium effect, as shown in Fig.\ref{fig:3}(b). The original lightred and lightblue spin textures move $\Delta k$ in the same direction, resulting in the red and blue spin textures. In contrast, for the conventional Rashba bands, the corresponding results are shown in Fig.\ref{fig:3}(a), the original lightred and lightblue spin textures move in opposite directions by $\Delta k$.
	
	In this section, we calculate the spin galvanic effect of the model, considering both potential impurity scattering and magnetic impurity scattering cases, respectively.
	
\subsection{The potential impurity scattering case}
	
	According to the semi-classical Boltzmann transport theory \cite{IngridMertig1999}, the shift of the Fermi circles in Fig.\ref{fig:3}(b) is equivalent to applying a homogeneous electrostatic field $\mathbf{E}$, which generates a directional current and causes the distribution function $f_k$ to deviate from the equilibrium distribution function $f_k^0$. Under the zero temperature limit and relaxation time approximation, it can be expresssed as
	\begin{align}
		f_k=f_k^0-|e|\tau_k \mathbf{v}_k \cdot \mathbf{E} \delta (E_k-E_F)
	\end{align}
	 The spin polarization $\langle \mathbf{S} \rangle$ is expressed as
	\begin{align}
		\langle \mathbf{S} \rangle = \sum_k \mathbf{S} (f_k-f_k^0)
	\end{align}
	where $e$ is the elementary charge, $\tau_k$ is the momentum relaxation time, and $\mathbf{v}_k$ is the group velocity.
	
	\begin{figure*}[t]
		\includegraphics[width=0.85 \linewidth]{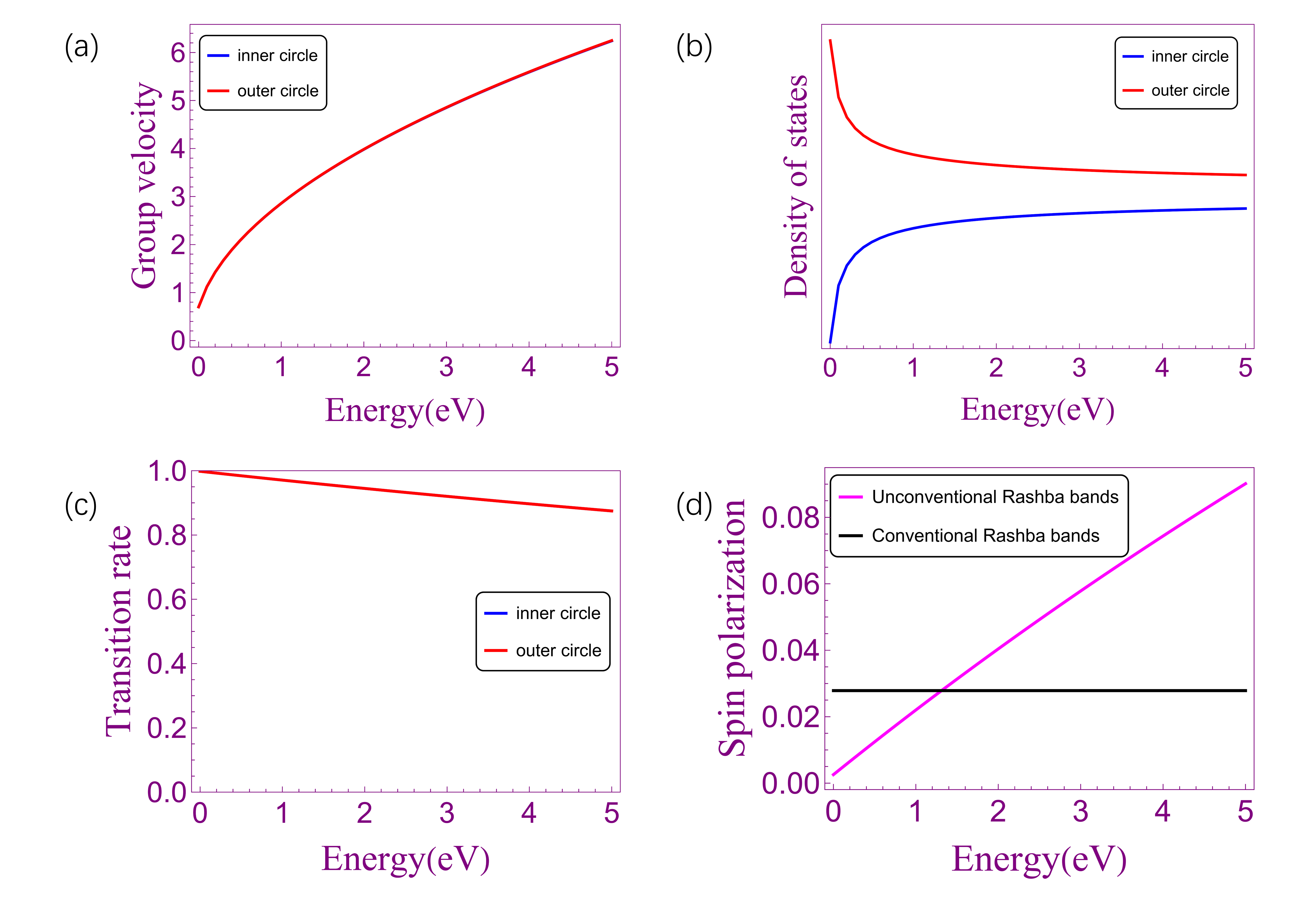}
		\caption{\label{fig:4}(a) Group velocity and (b) density of states of the inner and outer circles. (c) The inter-band scattering factor $|\langle \psi_{-\eta} (k,\theta) | \psi_{\eta} (k,\theta') |^2$ as a function of Fermi energy. (d) The total spin polarization of conventional and unconventional Rashba bands, where we adopt the same parameter of Rashba coupling $\lambda_{R}$.}
	\end{figure*}
	
	We choose the first and second bands in the $\mathbf{k} \cdot \mathbf{p}$ model for calculation, and represent them as
	\begin{align}
		E_\eta = k_{\eta}^2t + \eta k_{\eta}\lambda_R - \sqrt{\lambda^2+k_{\eta}^2\lambda_R^2} 
	\end{align}
	where $\eta=\pm$ represent the inner and outer bands, i.e., the blue and red bands, respectively. Correspondingly, the group velocity is
	\begin{align}
		v_{\eta}(k_{\eta})&=\frac{1}{\hbar}\frac{\partial}{\partial k_{\eta}} E_{\eta} =\frac{-k_{\eta}\lambda_R^2}{\hbar \sqrt{\lambda^2+k_{\eta}^2\lambda_R^2}}+\frac{2k_{\eta}t +\eta \lambda_R}{\hbar} 
	\end{align}
	In the $\mathbf{k} \cdot \mathbf{p}$ model, for a 2D isotropy system, the density of states is
	\begin{align}
		D(E_F)=\sum_k \delta(E_k-E_F)=\frac{A}{2\pi\hbar}\frac{k_F}{v_F}
	\end{align}
	where $k_F$ is the radius of Fermi circle at energy $E_F$, $v_F$ is the corresponding Fermi velocity, i.e., group velocity, $A$ is the area of the unit cell. The group velocity and density of states are shown in Fig.\ref{fig:4}(a) and (b). The group velocities of the inner and outer circles are nearly identical, indicating that the slopes of the two bands are almost identical. Regarding the density of states of the inner and outer circles, they exhibit exponential increase and decrease in the low-energy range (approximately $0-1.3 \ \text{eV}$), after which they become nearly parallel. This suggests that the distance between the two bands remains constant as $\mathbf{k}$ increases during the low-energy range.
	
	Now let's take a look at the relaxation time $\tau_F^\eta$. We have
	\begin{align}
		(\tau_F^\eta)^{-1}=\sum_{\eta'} P_{k_F^{\eta'} \leftarrow k_F^\eta}
	\end{align}
	which includes both intra-band and inter-band scattering.
	\begin{align}
		P_{k'\leftarrow k} = \frac{2\pi}{\hbar}N_{im}|T_{k'\leftarrow k}|^2\delta(E_{k'}-E_k)
	\end{align}
	Here, $P_{k'\leftarrow k}$ represents the probability of scattering per unit time. $N_{im}$ denotes the number of randomly distributed $\delta$-scatters in the system at various locations $\mathbf{r}_i$. Under the Born approximation, $T_{k'\leftarrow k}=\langle k' | \Delta V | k \rangle$, where $\Delta V$ represents the scattering potential. In this case, we consider short-range s-wave scattering with $\Delta V = \sum_i V_0 \delta(r-R_i)$. The intra-band and inter-band transition rates are related to the inner product of the corresponding wave functions, which is
	\begin{equation}
		\begin{split}
			& |\langle \psi_\eta (k,\theta) | \psi_\eta (k,\theta') |^2 = \frac{1}{2}[1+cos(\theta'-\theta)] \\
			& |\langle \psi_{\eta} (k,\theta) | \psi_{-\eta} (k,\theta') |^2 = \\
			& \qquad g(k_{\eta},k_{-\eta}) \frac{1}{2}[1-cos(\theta'-\theta)] \delta(E_{\eta}-E_{-\eta})
		\end{split}
	\end{equation}
	with
	\begin{equation}
		\begin{split}
			g(k_{\eta},k_{-\eta}) &= \frac{\lambda^2-k_{-\eta}k_{\eta}\lambda_R^2}{2\sqrt{(\lambda^2+k_{-\eta}^2\lambda_R^2)(\lambda^2+k_{\eta}^2\lambda_R^2)}} + \frac{1}{2}
		\end{split}
	\end{equation}
	At the same time, we can see that
	\begin{align}
		g(k_+,k_-)=g(k_-,k_+)
	\end{align}
	We can obtain the inter-band scattering factor $|\langle \psi_{-\eta} (k,\theta) | \psi_{\eta} (k,\theta') |^2$ as a function of Fermi energy, as shown in Fig.\ref{fig:4}(c). It monotonically decreases with increasing Fermi energy.
	
	Next, we need to calculate momentum relaxation time at the Fermi level. For isotropy system, we have 
	\begin{align}
		\sum_{\mathbf{k}}p(\theta)F(k)\delta(E_k-E_F) = F(k_F)D(E_F)\frac{1}{2\pi} \int_0^{2\pi}p(\theta) d\theta
	\end{align}	
	The momentum relaxation time can be expressed as
	\begin{align} \label{eq28}
		\tau_F^{\eta}&=\frac{1}{\frac{2\pi}{\hbar}N_{im}V_0^2[D(E_F^{\eta})+g(k_F^{-\eta},k_F^{\eta})D(E_F^{-\eta})]}
	\end{align}
	Once the group velocity, momentum relaxation time, and spin texture are obtained, the spin polarization $\langle \mathbf{S} \rangle$ can be expressed as
	\begin{equation} \label{eq30}
		\begin{split}
			\langle \mathbf{S} \rangle 
			&= -|e| \sum_k \mathbf{S}_k(\tau_k \mathbf{v}_k \cdot \mathbf{E}) \delta(E_k - E_F) \\
			&= \frac{|e|A}{2\pi\hbar}\sum_\eta I^\eta(k_F^\eta)\tau_F^\eta k_F^\eta (\hat{\mathbf{v}}_F^\eta \cdot \hat{\mathbf{k}}_F^\eta)(\hat{\mathbf{z}}\times \mathbf{E}) 
		\end{split} 
	\end{equation}
	where
	\begin{align*}
		I^\eta(k)=\frac{k^\eta \lambda_R}{\sqrt{{k^\eta}^2\lambda_R^2+\lambda^2}}
	\end{align*}
	Through Fq.(\ref{eq30}), we obtain the total spin polarization as shown in Fig.\ref{fig:4}(d). It can be seen that the spin polarization of unconventional Rashba bands and conventional Rashba bands intersect at $1.3 \ \text{eV}$, indicating a competition between them. The spin polarization of unconventional Rashba bands increases linearly, which is not conducive to its application at high energies.
	
\begin{figure}[t]
		\includegraphics[width=0.9 \linewidth]{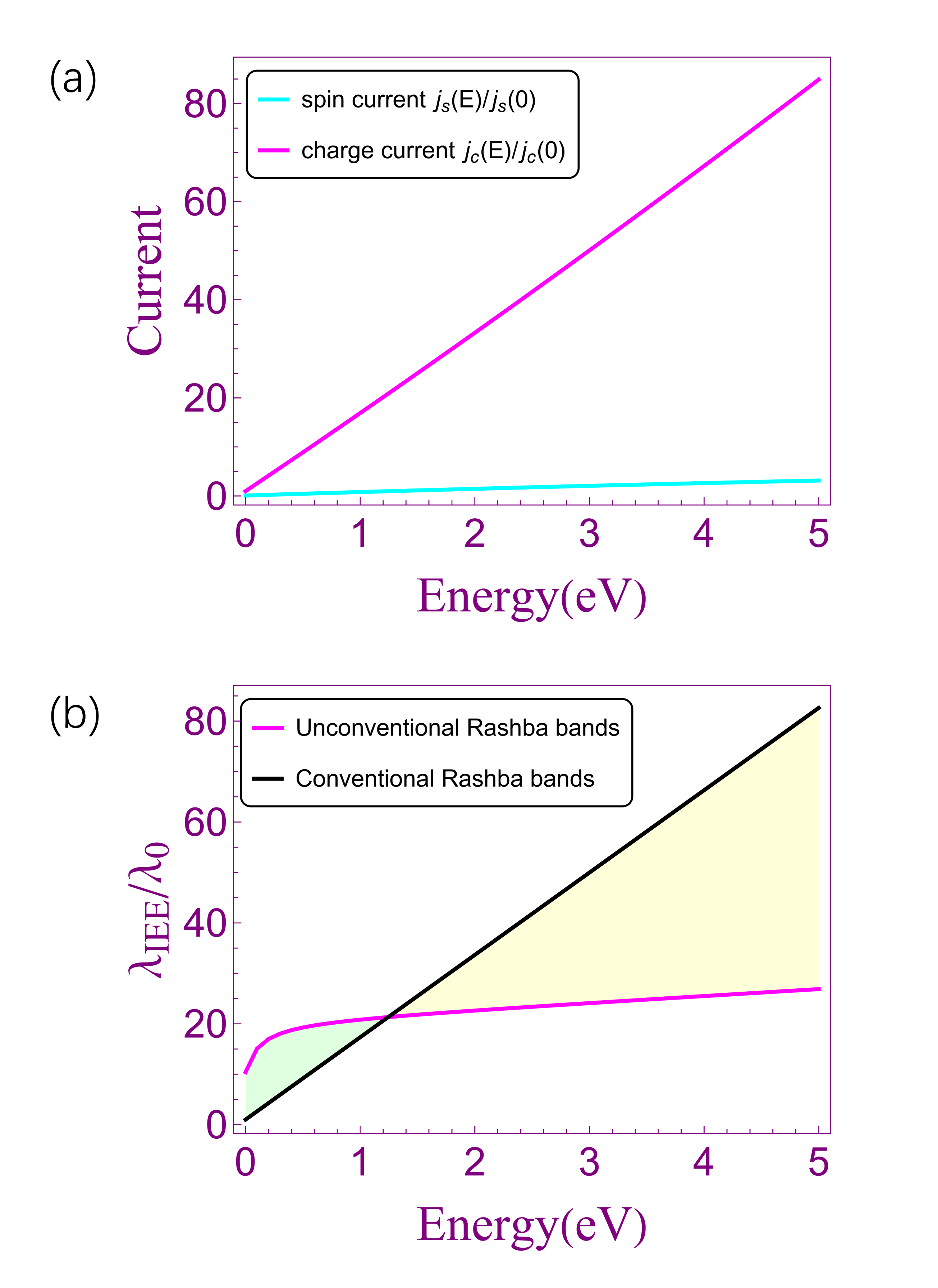}
		\caption{\label{fig:5}(a) The relative spin current $j_s(E_F)/j_s(0)$ and the relative charge current $j_c(E_F)/j_c(0)$ as the function as the Fermi energy for the unconventional Rashba bands. (b) The spin-to-charge conversion efficiency $\lambda_{IEE}$ of conventional and unconventional Rashba bands as the function as the Fermi energy.}
	\end{figure}

	We focus on the spin-to-charge conversion, spin current density is represented as
	\begin{align}\label{eq31}
		\mathbf{j}_s=\frac{e\langle S \rangle}{\tau} = \sum_\eta \frac{e\langle S \rangle ^\eta }{\tau^\eta} \hat{\mathbf{z}}
	\end{align}
	And charge current density can be derived as
	\begin{align}\label{eq32}
		\mathbf{j}_c=e \sum_k \mathbf{v}_k (f_k - f_k^0) = \sum_\eta e^2 \mathbf{v}_F^\eta (\tau_F^\eta \mathbf{v}_F^\eta \cdot \mathbf{E}) D(E_F^\eta)
	\end{align}
	where $\eta=\pm$ represent inner and outer Fermi circle. Note that we will take $j_s=|\mathbf{j}_s|$ and $j_c=|\mathbf{j}_c|$ in the subsequent calculations. Therefore, we can obtain the relative spin current $j_s(E_F)/j_s(0)$ and the relative charge current $j_c(E_F)/j_c(0)$ as the function as the Fermi energy for the unconventional Rashba systems, as shown in Fig.\ref{fig:5}(a). It can be observed that both charge current and spin current increase with increasing Fermi energy. However, as the Fermi energy increases, the charge current at the same Fermi energy is significantly greater than the spin current.
	
	The conversion efficiency $\lambda_{IEE}$ from spin current to charge current can be written as
	\begin{align}
		\lambda_{IEE}=\frac{{j}_c}{{j}_s}=\frac{\sum_\eta k_F^\eta v_F^\eta \tau_F^\eta}{\sum_\eta I^\eta k_F^\eta (\hat{\mathbf{v}}_F^\eta \cdot \hat{\mathbf{k}}_F^\eta)}
	\end{align}
	The spin-to-charge conversion efficiency $\lambda_{IEE}$ of unconventional Rashba bands as the function as the Fermi energy, as shown in Fig.\ref{fig:5}(b). Here, $\lambda_{IEE}$ is in unit of $\lambda_{0}$, which is the $\lambda_{IEE}$ of conventional Rashba bands at $E_F = 0$. Between $0-1.3 \ \text{eV}$, the $\lambda_{IEE}$ of unconventional Rashba bands surpasses that of conventional Rashba bands, as depicted in the lightgreen region in Fig.\ref{fig:5}(b). Beyond $1.3 \ \text{eV}$, represented by the lightyellow region in the Fig.\ref{fig:5}(b), the conventional Rashba bands exhibits superior performance.
	
	For the conventional Rashba bands mentioned earlier, we consider
	\begin{align}
		E_\eta &= k_\eta^2 t + \eta k_\eta \lambda_{R} - \lambda
	\end{align}
	At this point, the spin texture satisfies
	\begin{align}
		\mathbf{S}_{\mathbf{k}_F^-}^- = \mathbf{S}_{\mathbf{k}_F^+}^+
	\end{align}
	By incorporating the above formula, we can calculate the total spin polarization and $\lambda_{IEE}$. 
	
	For Fig.\ref{fig:5}(b), combined with Fig.\ref{fig:4}(d), we see that unconventional Rashba bands is better than conventional Rashba bands between $0-1.3 \ \text{eV}$. That is to say, at low energy, the Fermi energy is closer to the intersection of the two coupled bands, and unconventional Rashba bands is advantageous as it facilitates the conversion of spin current to charge current.
	
	\begin{figure*}[t]
		\includegraphics[width=0.9 \linewidth]{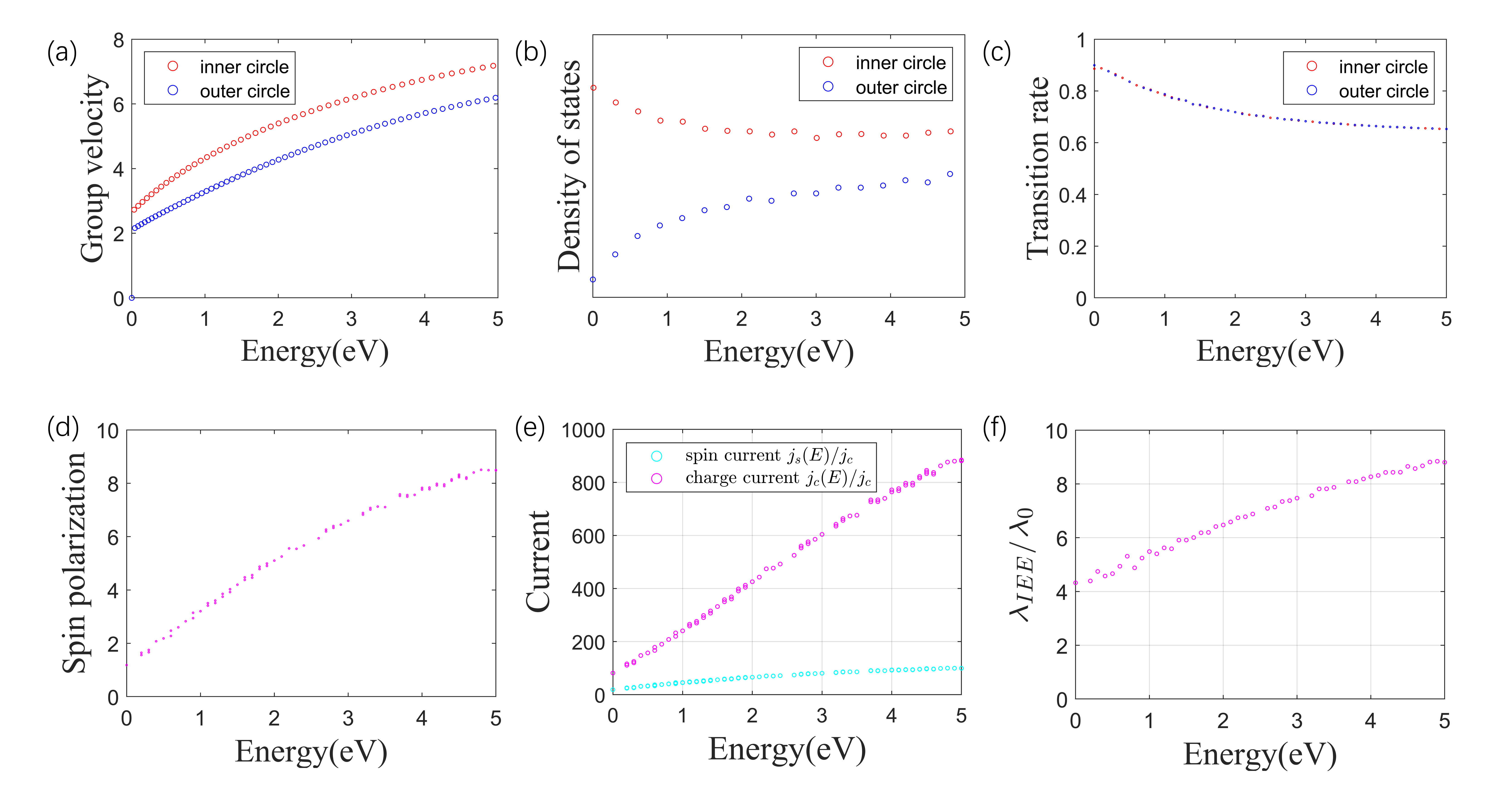}
		\caption{\label{fig:6}(a) Group velocity and (b) density of states of inner and outer circles. (c) The inter-band scattering factor $|\langle \psi_{-\eta} (k,\theta) | \psi_{\eta} (k,\theta') |^2$ as a function of Fermi energy. (d) The total spin polarization of the unconventional Rashba bands. (e) The relative spin current $j_s(E_F)/j_s(0)$ and the relative charge current $j_c(E_F)/j_c(0)$ as the function as the Fermi energy for the unconventional Rashba bands. (f) The spin-to-charge conversion efficiency $\lambda_{IEE}$ of the unconventional Rashba bands as the function as the Fermi energy.}
	\end{figure*}
	
	By analytically solving the $\mathbf{k} \cdot \mathbf{p}$ model, we can directly compute spin polarization, spin current, and charge current, and $\lambda_{IEE}$. More precisely, we perform numerical calculations in the TB model of the hexagonal lattice, as shown in Fig.\ref{fig:1}(b). We select the first and second bands (i.e., the red and blue bands) for calculation.
	
	For the group velocity and density of states of the TB model, as shown in Fig.\ref{fig:6}(a) and (b), its trend is consistent with that of the $\mathbf{k} \cdot \mathbf{p}$ model. The group velocity reflects the variation of energy in the k-space. From the group velocity of the TB model, there is a slight splitting of the inner and outer band group velocities, which reflects the more complex bands of the TB model and the enhanced coupling in the k-space. Although the group velocity has split, the curve of the density of states is roughly consistent with Fig.\ref{fig:4}(b). This indicates that the state of energy does not change much on the same equipotential surface. In Fig.\ref{fig:6}(c), we see that the intra-band scattering and inter-band scattering of the two coupled bands decrease with the increase of the Fermi level. This is consistent with Fig.\ref{fig:4}(c). The curve in Fig.\ref{fig:4}(c) shows a continuous decreasing trend as well. Through this comparison, we can say that the $\mathbf{k} \cdot \mathbf{p}$ model well reflects the TB model in the low energy range (around $0-5 \ \text{eV}$). Similarly, by comparing Fig.\ref{fig:6}(d) and Fig.\ref{fig:4}(d), the overall spin polarization also shows the same trend. The spin current and charge current require our attention, as $\lambda_{IEE}={j}_c/{j}_s$ closely affects $\lambda_{IEE}$. Through the previous judgment, we found that in the low energy range, the trend of $\lambda_{IEE}$ change is consistent with that obtained by the previous $\mathbf{k} \cdot \mathbf{p}$ model, as shown in Fig.\ref{fig:6}(e) and (f). There is no comparison between unconventional Rashba bands and conventional Rashba bands here because the previous conventional Rashba bands used the $\mathbf{k} \cdot \mathbf{p}$ model, which is meaningless to compare.
	
\subsection{The magnetic impurity scattering case}
	
	The above transport calculations are taking into account the potential impurity scattering case. Now, we aim to calculate the scattering rate $1/\tau$ in the presence of magnetic impurities. The unperturbed Hamiltonian Eq.(\ref{eq13}) is written in the bases of spin and orbits, while the scattering Hamiltonian reads
	\begin{align}
		&H_{\text {int }} =J\int\mathrm{ d} \textbf{x} \psi_{\sigma \mu}^{\dagger}(\textbf{x}) V_{\sigma \sigma^{\prime}}^{\mu \mu^{\prime}}(\textbf{x}) \psi_{\sigma \mu^{\prime}}(\textbf{x})\\
		&V_{\sigma \sigma^{\prime}}^{\mu \mu^{\prime}}(\textbf{x})=\bm{\sigma}_{\sigma \sigma^{\prime}} \cdot \textbf{S}_{d}(\textbf{x}) \delta_{\mu \mu^{\prime}} \\
		&\textbf{S}_{d}(\textbf{x})=\textbf{S}_{d} \textstyle\sum_{j} \delta\left(\textbf{x}-\textbf{P}_{j}\right)
	\end{align}
	where  $\mu (\mu^{\prime})=d_{x^2-y^2} \ \text{or} \ d_{xy}$ orbitals and $\sigma (\sigma^{\prime})$ stand for orbital and spin degrees of freedom respectively, and $\textbf{P}_{j}$ are positions of impurities. The Dyson equation of electron's Green's function takes the standard form $\mathcal{G} (b, a)=\mathcal{G}^{0}(b, a)+\int \mathrm{d}1\mathrm{d}1^{\prime} \mathcal{G}^{0}(b, 1)V\left(1,1^{\prime}\right) \mathcal{G}\left(1^{\prime}, a\right)$, where we have used the abstract indices labelling all degrees of freedom. For the Matsubara Green's function defined as $\mathcal{G}_{\sigma_{b}\sigma_{a}}^{\mu_{b}\mu_{a}}\left ( \textbf{x}_{b},\textbf{x}_{a}; \tau_{b}-\tau_{a} \right ) \equiv -\left \langle  T_{\tau}\psi_{\sigma_{b}\mu_{b}}\left ( \text{x}_{b},\tau_{b} \right )\psi^{\dagger}_{\sigma_{a}\mu_{a}}\left ( \text{x}_{a},\tau_{a} \right ) \right \rangle $, the Dyson equation in frequency domain can be written specifically
	\begin{widetext}
	\begin{align}
		\mathcal{G}_{\sigma_{b}\sigma_{a}}^{\mu_{b}\mu_{a}}\left ( \textbf{x}_{b},\textbf{x}_{a}; ik_{n} \right ) &= \mathcal{G}_{\sigma_{b}\sigma_{a}}^{0,\mu_{b}\mu_{a}}\left ( \textbf{x}_{b}-\textbf{x}_{a}; ik_{n} \right ) \nonumber\\
		&+\operatorname{Tr} \sum_{j} \sum_{\substack{\sigma_{1}, \sigma_{1}^{\prime}, \mu_{1}}} \mathcal{G}_{\sigma_{b} \sigma_{1}}^{0 \mu_{b} \mu_{1}}\left(\textbf{x}_{b}-\textbf{P}_{j}, i k_{n}\right) \bm{\sigma}_{\sigma_{1}, \sigma_{1}^{\prime}} \cdot \textbf{S}_{d} \ \mathcal{G}_{\sigma_{1}^{\prime} \sigma_{a}}^{\mu_{1} \mu_{a}}\left(\textbf{P}_{j}, \textbf{x}_{a ;} ; ik_{n}\right)
	\end{align}
	\end{widetext}
	where the trace operator acts on the spin space of isolated impurity. In virtue of the Dyson equation, we could expand the full Green's function in the power of coupling constant $J$. The n-order term reads
	\begin{widetext}
	\begin{align}
		\mathcal{G}_{\sigma_{b}\sigma_{a}}^{(n),\mu_{b}\mu_{a}}&\left ( \textbf{x}_{b},\textbf{x}_{a}; ik_{n} \right ) = \operatorname{Tr} \sum_{j_{1} \cdots j_{n}}\sum_{\substack{ \sigma_{1} \cdots \sigma_{n} \\ \sigma_{1}^{\prime} \cdots \sigma_{n}^{\prime} \\ \mu_{1} \cdots \mu_{n}}}    \mathcal{G}_{\sigma_{b} \sigma_{1}}^{0 \mu_{b} \mu_{1}}\left(\textbf{x}_{b}-\textbf{P}_{j_{1}}, ik_{n}\right) \bm{\sigma}_{\sigma_{1} \sigma_{1}^{\prime}} \cdot \textbf{S}_{d}\ \mathcal{G}_{\sigma_{1}^{\prime} \sigma_{2}}^{0 \mu_{1} \mu_{2}}\left(\textbf{P}_{j_{1}}-\textbf{P}_{j_{2}}, ik_{n}\right)\nonumber\\
		&\times \bm{\sigma}_{\sigma_{2} \sigma_{2}^{\prime}} \cdot \textbf{S}_{d}\ \mathcal{G}_{\sigma_{2^{\prime}} \sigma_{3}}^{0 \mu_{2} \mu_{3}}\left(\textbf{P}_{j_{2}}-\textbf{P}_{j_{3}}, ik_{n}\right) \times \cdots \times \bm{\sigma}_{\sigma_{n} \sigma_{n}^{\prime}} \cdot \textbf{S}_{d}\ \mathcal{G}_{\sigma_{n}^{\prime} \sigma_{a}}^{0 \mu_{n} \mu_{a}}\left(\textbf{P}_{j_{n}}-\textbf{x}_{a}, ik_{n}\right)
	\end{align}
	\end{widetext}
	Now we get down to transform into the band presentation $\mathcal{G}_{\sigma \sigma^{\prime}}^{\mu \mu^{\prime}}\left(\textbf{x}, \textbf{x}^{\prime}\right) \Rightarrow  \mathcal{G}_{\nu\nu^{\prime}}\left(\textbf{k}, \textbf{k}^{\prime}\right)$, where $\nu(\nu^{\prime})=1,\cdots, 4$ is band index. Using a simple relation $\langle\nu \textbf{k} \, | \, \mu \sigma \textbf{x}\rangle =\frac{1}{V}e^{i \textbf{k}\cdot\textbf{x}}\langle\nu \textbf{k} \, |\,  \mu \sigma \textbf{k}\rangle
	$, we obtain the followig representation transformation,
	\begin{widetext}
	\begin{align}
		\mathcal{G}_{\sigma \sigma^{\prime}}^{\mu \mu^{\prime}}\left(\textbf{x}, \textbf{x}^{\prime}\right) & =\frac{1}{V^{2}}\sum_{\nu \nu^{\prime}} \sum_{\textbf{k} \textbf{k}^{\prime}}\langle\mu \sigma \textbf{x}\,|\, \nu \textbf{k}\rangle\mathcal{ G}_{\nu \nu^{\prime}}\left(\textbf{k}, \textbf{k}^{\prime}\right)\left\langle\nu^{\prime} \textbf{k}^{\prime}\,|\, \mu^{\prime} \sigma^{\prime} \textbf{x}^{\prime}\right\rangle \nonumber \\
		& =\frac{1}{V^{2}}\sum_{\nu \nu^{\prime}} \sum_{\textbf{k} \textbf{k}^{\prime}}\langle\mu \sigma \textbf{k}\,|\, \nu \textbf{k}\rangle \mathcal{G}_{\nu \nu^{\prime}}\left(\textbf{k}, \textbf{k}^{\prime}\right)\left\langle\nu^{\prime} \textbf{k}^{\prime}\,|\, \mu^{\prime} \sigma^{\prime} \textbf{k}^{\prime}\right\rangle e^{i \textbf{k}^{\prime} \cdot \textbf{x}^{\prime}-i \textbf{k} \cdot \textbf{x}}
	\end{align}
	\end{widetext}
	Hence the n-order Green's function in the bases of energy bands takes the form
	\begin{widetext}
	\begin{align}
		\mathcal{G}_{\nu_{b} \nu_{a}}^{(n)}\left(\textbf{k}_{b}, \textbf{k}_{a}\right)&=J^{n}\sum_{j_{1} \cdots j_{n}}\sum_{\substack{\sigma_{1}\cdots\sigma_{n} \\ \sigma_{1}^{\prime} \cdots \sigma_{n}^{\prime} \\ \mu_{1} \cdots \mu_{n}}} \frac{1}{V^{n-1}} \sum_{\textbf{k}_{1} \cdots \textbf{k}_{n-1}} \sum_{\nu_{1} \cdots \nu_{n-1}} \left\langle\nu_{b} \textbf{k}_{b} \,|\, \sigma_{n} \mu_{n} \textbf{k}_{b}\right\rangle\left\langle\sigma_{n}^{\prime} \mu_{n} \textbf{k}_{n-1} \mid \nu_{n-1} \textbf{k}_{n-1}\right\rangle  \nonumber\\
		&\times\left\langle\nu_{n-1} \textbf{k}_{n-1} \,|\, \sigma_{n-1} \mu_{n-1} \textbf{k}_{n-1}\right\rangle\left\langle\sigma_{n-1}^{\prime} \mu_{n-1} \textbf{k}_{n-2} \,|\, \nu_{n-2} \textbf{k}_{n-2}\right\rangle \nonumber\\
		&\times \cdots \times\left\langle\nu_{1} \textbf{k}_{1} \,|\, \sigma_{1} \mu_{1} \textbf{k}_{1}\right\rangle\left\langle\sigma_{1}^{\prime} \mu_{1} \textbf{k}_{a} \,|\, \nu_{a} \textbf{k}_{a}\right\rangle\nonumber\\
		&\times \mathcal{G}_{\nu_{b}}^{0}\left(\textbf{k}_{b}\right) \mathcal{G}_{\nu_{n-1}}^{0}\left(\textbf{k}_{n-1}\right) \mathcal{G}_{\nu_{n-2}}^{0}\left(\textbf{k}_{n-2}\right) \times\cdots \times\mathcal{G}_{\nu_{1}}^{0}\left(\textbf{k}_{1}\right) \mathcal{G}_{\nu_{a}}^{0}\left(\textbf{k}_{a}\right)\nonumber\\
		&\times e^{-i\left(\textbf{k}_{b}-\textbf{k}_{n-1}\right)\cdot\textbf{P}_{j_{n}}} e^{-i\left(\textbf{k}_{n-1}-\textbf{k}_{n-2}\right) \cdot \textbf{P}_{j_{n-1}}} \times \cdots \times e^{-i\left(\textbf{k}_{1}-\textbf{k}_{a}\right) \cdot \textbf{P}_{j_{1}}}\nonumber\\
		&\times \operatorname{Tr}\left[\left(\bm{\sigma}_{\sigma_{n} \sigma_{n}^{\prime}} \cdot \textbf{S}_{d}\right)\left(\bm{\sigma}_{\sigma_{n-1} \sigma_{n-1}^{\prime}} \textbf{S}_{d}\right) \times \cdots \times\left(\bm{\sigma}_{\sigma_{1}, \sigma_{1}^{\prime}} \cdot \textbf{S}_{d}\right)\right]\label{green's function}
	\end{align}
	\end{widetext}
	where the frequency $ik_{n}$ is implied and $\mathcal{G}_{\nu}^{0}\left(\textbf{k}, i{k}_{n}\right)=1/(ik_n-\epsilon_{\textbf{k}}^{\nu})$. Note that $\epsilon_{\textbf{k}}^{\nu}$ here is the eigenvalue of Eq.(\ref{eq13}), and $\nu$ is band index. One can perform the position average over all impurity configurations since these impurities are random and uncorrelated, which gives rise to a momentum-conserving Green's function. After the impurity-average procedure, $\frac{1}{V} \left \langle \mathcal{G}\left ( \textbf{k}_{b},\textbf{k}_{a} \right )  \right \rangle_{imp}\equiv\delta_{\textbf{k}_{b},\textbf{k}_{a}}\frac{1}{V}\int \mathrm{d}\textbf{P}_{1}\cdots\frac{1}{V}\int \mathrm{d}\textbf{P}_{N_{imp}}\mathcal{G}\left ( \textbf{k}_{b},\textbf{k}_{a} \right ) $, we get the Green's function with the same outgoing and incoming momentum which can be presented graphically.
	\begin{widetext}
	\begin{align}
		\left \langle \mathcal{G}_{\nu_{b} \nu_{a}}^{(n)}\left(\textbf{k}\right) \right \rangle_{imp} &=J^{n}\textstyle{\sum_{p=1}^{n}}\textstyle{\sum_{All\ diagrams\ with\ p\ impurities}} \sum_{\nu_{1} \cdots \nu_{n-1}}\sum_{\mu_{1} \cdots \mu_{n}}\sum_{\substack{\sigma_{1}\cdots\sigma_{n} \\ \sigma_{1}^{\prime} \cdots \sigma_{n}^{\prime}}} (n_{imp})^{p} \nonumber \\
		&\times\left\langle\nu_{b} \textbf{k} \,|\, \sigma_{n} \mu_{n} \textbf{k}\right\rangle\left\langle\sigma_{n}^{\prime} \mu_{n} \textbf{k}\mid \nu_{n-1} \textbf{k}\right\rangle  \cdots \left\langle\nu_{1} \textbf{k} \,|\, \sigma_{1} \mu_{1} \textbf{k}\right\rangle\left\langle\sigma_{1}^{\prime} \mu_{1} \textbf{k} \,|\, \nu_{a} \textbf{k}\right\rangle\nonumber \\
		&\times \mathcal{G}_{\nu_{b}}^{0}\left(\textbf{k}\right) \mathcal{G}_{\nu_{n-1}}^{0}\left(\textbf{k}\right)  \cdots \mathcal{G}_{\nu_{1}}^{0}\left(\textbf{k}\right) \mathcal{G}_{\nu_{a}}^{0}\left(\textbf{k}\right)\nonumber \\
		&\times \operatorname{Tr}\left[\left(\bm{\sigma}_{\sigma_{n} \sigma_{n}^{\prime}} \cdot \textbf{S}_{d}\right)\left(\bm{\sigma}_{\sigma_{n-1} \sigma_{n-1}^{\prime}} \textbf{S}_{d}\right)  \cdots \left(\bm{\sigma}_{\sigma_{1}, \sigma_{1}^{\prime}} \cdot \textbf{S}_{d}\right)\right]
	\end{align}
	\end{widetext}

	\begin{figure}[t]
		\includegraphics[width=0.9 \linewidth]{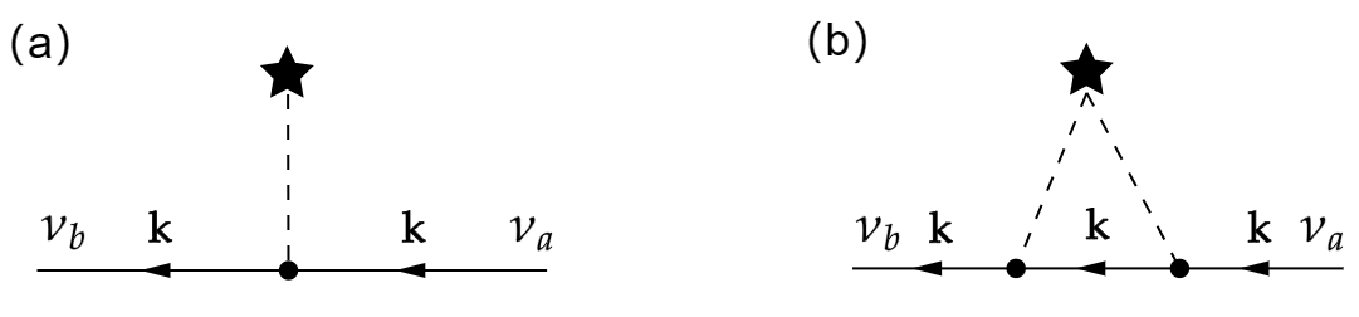}
		\caption{\label{fig:7}\textbf{Self-energy:} (a) lowest-order approximation; (b) first-order Born approximation.}
	\end{figure}

	Notice that all the exponentials in Eq.(\ref{green's function}) are gone and all the intermediate processes are of the same momentum since the momentum conservation is required  and a constant coupling strength $J$ is presupposed.
	We now take two well-known approximations, namely the lowest-order approximation (LOA) and the first-order Born approximation (FBA), shown in Fig.(\ref{fig:7}). LOA diagram which is proportional to $\operatorname{Tr}\left(\bm{\sigma}_{\sigma \sigma} \cdot \textbf{S}_{d}\right)$ is clearly zero owing to the traceless feature of impurity spin operator $\textbf{S}_{d}$, while FBA diagram has non-zero contribution
	\begin{widetext}
	\begin{align}
		{\LARGE\Sigma }  _{\nu_{b}\nu_{a}}^{FBA}\left(\textbf{k}, ik_{n}\right)&=
		n_{i m p}J^{2} \sum_{\substack{\sigma_{1} \sigma_{2} \\ \sigma_{1}^{\prime} \sigma_{2}^{\prime}}}\sum_{\mu_{1}\mu_{2},\nu}\left\langle\nu_{b} \textbf{k} \mid \sigma_{2} \mu_{2} \textbf{k}\right\rangle\left\langle\sigma_{2}^{\prime} \mu_{2} \textbf{k} \mid \nu \textbf{k}\right\rangle\left\langle\nu \textbf{k} \mid \sigma_{1} \mu_{1} \textbf{k}\right\rangle\left\langle\sigma_{1}^{\prime} \mu_{1} \textbf{k} \mid \nu_{a} \textbf{k}\right\rangle\nonumber\\
		&\times\operatorname{Tr}\left[\left(\bm{\sigma}_{\sigma_{2} \sigma_{2}^{\prime}} \cdot \textbf{S}_{d}\right)\left(\bm{\sigma}_{\sigma_{1} \sigma_{1}^{\prime}} \cdot \textbf{S}_{d}\right)\right] \mathcal{G}_{\nu}^{0}\left(\textbf{k}, ik_{n}\right)\label{FBA self-energy}
	\end{align}
	
	\end{widetext}
	where all inner-products can be evaluated specifically considering that we have already diagonalized the unperturbed Hamiltonian. We then could easily obtain the retarded self-energy by implementing the analytic continuation $ik_{n}\to \omega+i0^{+}$. Suppose that the Fermi energy is located in the lower two bands and that $\lambda\gg\lambda_{R}k_{F}$, we calculate all inner-products within these bands, as shown in Table \ref{inner-products}, where $\theta=\arctan(k_{y}/k_{x})$ and these expressions are expanded to the first order in $\lambda_{R}k/\lambda$. Taking the imaginary part of the self-energy and averaging it over the Fermi surface by integrating momentum, we find an averaged scattering rate $\tau_{i\gets j}^{-1}\equiv\frac{1}{2\pi}\textstyle{\sum_{\textbf{k}}}(\tau_{\textbf{k}}^{i\gets j})^{-1}=-\frac{1}{2}\textstyle{\sum_{\textbf{k}}}Im{\LARGE\Sigma}^{R}_{ij}(\textbf{k},\omega=E_{F})$, where $i(j)=1,2$ (corresponds to $-,+$ in the index convention above), refering to the lower two bands. Using the identity that, for an arbitrary function $f(k,\theta)$, $\textstyle{\sum_{\textbf{k}}}f(k,\theta)\delta(E_{F}-\epsilon_{\textbf{k}}^{i})=(Sk_{F}^{i}/4\pi^{2}v_{F}^{i})\int_{0}^{2\pi}\mathrm{d}\theta f(k_{F}^{i},\theta)$, where $S$, $v_{F}^{i}$ and $k_{F}^{i}$ are the area of sample, Fermi velocity and Fermi wave vector of i'th band, respectively, we obtain the following results for the total scattering rate of i'th band
	\begin{align}
		\tau_{1}^{-1} & =\tau_{1 \gets 1}^{-1}+\overbrace{\tau_{2 \gets 1}^{-1}}^{0}\nonumber\\
		& =4 \pi J^{2} n_{i m p}\left[D_{1}\left(E_{F}\right)\left(\frac{k_{F} \lambda_{R}}{\lambda}\right)^{2}+D_{2}\left(E_{F}\right)\right]  \label{eq44} \\
		\tau_{2}^{-1} & =\tau_{2 \gets 2}^{-1}+\overbrace{\tau_{1 \gets 2}^{-1}}^{0}\nonumber\\
		& =4 \pi J^{2} n_{i m p}\left[D_{2}\left(E_{F}\right)\left(\frac{k_{F} \lambda_{R}}{\lambda}\right)^{2}+D_{1}\left(E_{F}\right)\right]  \label{eq45}
	\end{align}
	where $D_{i}(E_{F})$ is the density of states at Fermi energy of i'th band. From the detailed calculation of Eq.(\ref{FBA self-energy}), we note that, due to the procedure of averaging over the Fermi surface, terms in the summation cancel each other out if $i\ne j$, ending up with a vanishing inter-band averaged scattering rate.
	
	\begin{table}[]
		\renewcommand\arraystretch{1.6}
		\centering
		\caption{Inner-products}
		\label{inner-products}
		\begin{tabular*}{\linewidth}{lcc}
			\hline\hline
			\qquad\qquad& $\left | 1,\textbf{k} \right \rangle$ \qquad\qquad& $\left | 2,\textbf{k} \right \rangle$ \\
			\hline
			$\left \langle d_{x^2-y^2},\uparrow ,\textbf{k} \right |$ \qquad\qquad& $-\sin\theta-i\cos\theta$ \qquad\qquad& $\sin\theta+i\cos\theta$ \\
			$ \left \langle d_{x^2-y^2},\downarrow ,\textbf{k} \right |$ \qquad\qquad& $-k\lambda_{R}/\lambda-i$ \qquad\qquad& $k\lambda_{R}/\lambda-i$ \\
			$\left \langle d_{xy},\uparrow ,\textbf{k} \right |$ \qquad\qquad& $-e^{-i\theta}(1-ik\lambda_{R}/\lambda)$ \qquad\qquad& $e^{-i\theta}(1+ik\lambda_{R}/\lambda)$ \\
			$\left \langle d_{xy},\downarrow ,\textbf{k}\right |$ \qquad\qquad& 1 \qquad\qquad& 1 \\
			\hline\hline
		\end{tabular*}
	\end{table}
	
	\begin{figure}[t]
		\includegraphics[width=0.9 \linewidth]{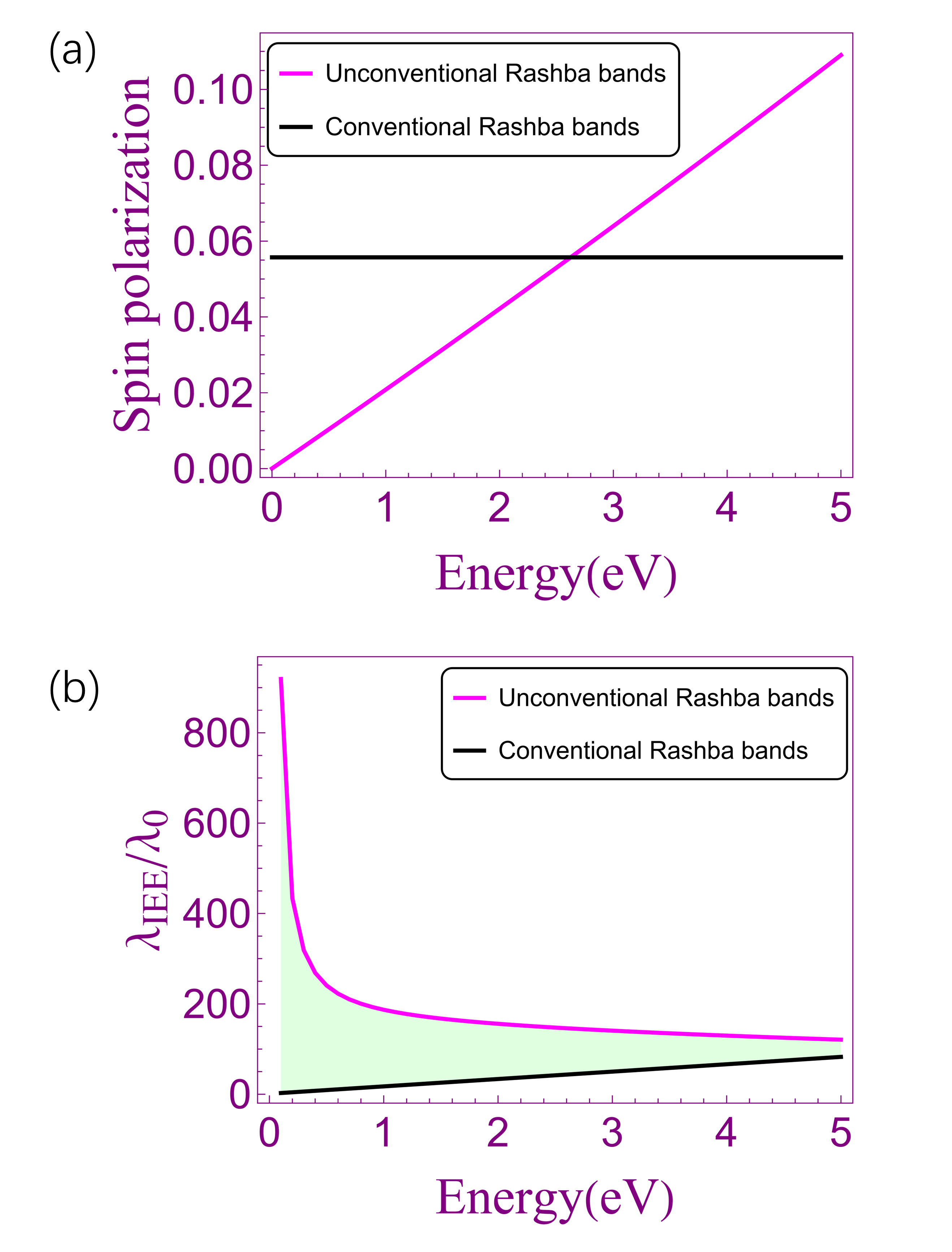}
		\caption{\label{fig:8}(a) The total spin polarization of conventional and unconventional Rashba bands. (b) The spin-to-charge conversion efficiency $\lambda_{IEE}$ of conventional and unconventional Rashba bands as the function as the Fermi energy.}
	\end{figure}
	
	We write Eq.(\ref{eq44}) and Eq.(\ref{eq45}) in the form of Eq.(\ref{eq28}), which is represented as
	\begin{align}
		\tau_F^{\eta}&=\frac{1}{4 \pi J^{2} n_{i m p}\left[D\left(E_{F}^{\eta}\right)\left(\frac{k_{F} \lambda_{R}}{\lambda}\right)^{2}+D\left(E_{F}^{-\eta}\right)\right]}
	\end{align}
	where $\eta=\pm$, corresponds to 2 and 1, respectively. It can be seen that in the magnetic impurity scattering case, the momentum relaxation time is different, so the total spin polarization and $\lambda_{IEE}$ are also different, as shown in Fig.\ref{fig:8}(a) and (b). We can observe that the spin polarization has a linear feature in a large energy range and the spin-to-charge conversion efficiency $\lambda_{IEE}$ significantly deviates from the potential-scattering situation. Fig.\ref{fig:8} (b) shows that $\lambda_{IEE}$ is greatly enhanced compared to potential scattering and exceeds the conventional Rashba SOC throughout the entire energy range, which exhibits the giant spin galvanic effect.
	
\section{CONCLUSIONS}
  In this paper, we constructed a four-band TB model based on two-dimensional hexagonal and square lattices, considering the $d_{x^2-y^2}$ and $d_{xy}$ orbitals of individual atoms, as well as nearest-neighbor hopping, on-site SOC, and nearest-neighbor Rashba SOC. Both TB models exhibit globally unconventional Rashba bands. Subsequently, we discussed the $\mathbf{k} \cdot \mathbf{p}$ model near the $\Gamma$ point and calculated the spin galvanic effect under potential impurity scattering. Comparison with numerical calculations from the TB model yielded nearly identical results, indicating the presence of a giant spin galvanic effect in this model. Furthermore, we explored the case of magnetic impurity scattering and found that the spin galvanic effect is significantly larger in this case, surpassing the effects observed in potential impurity scattering. Our findings highlight the existence of a giant spin galvanic effect in unconventional Rashba bands and provide important insights for further exploration in experimental materials.
  
\begin{acknowledgments}
This work was financially
supported by the National Key R\&D Program of China (Grants No.
2022YFA1403200), National Natural Science Foundation of
China (Grants No. 92265104, No. 12022413, No. 11674331), the Basic Research Program of the Chinese
Academy of Sciences Based on Major Scientific Infrastructures (Grant No. JZHKYPT-2021-08), the CASHIPS Director’s Fund (Grant No. BJPY2023A09), the \textquotedblleft
Strategic Priority Research Program (B)\textquotedblright\ of the Chinese
Academy of Sciences (Grant No. XDB33030100), and the Major Basic Program of Natural
Science Foundation of Shandong Province (Grant No. ZR2021ZD01). A portion of
this work was supported by the High Magnetic Field Laboratory of Anhui
Province, China.
\end{acknowledgments}

\end{document}